\documentclass[aps,prb,twocolumn,showpacs,groupedaddress, amsmath]{revtex4}

\usepackage{graphicx}
\usepackage{amssymb}

\begin{document}


\title{Observation of superspin-glass behavior in Fe$_{3}$O$_{4}$ nanoparticles}


\author{Masatsugu Suzuki}
\email[]{suzuki@binghamton.edu}
\affiliation{Department of Physics, State University of New York at Binghamton, Binghamton, New York 13902-6000}

\author{Sharbani I. Fullem}
\affiliation{Department of Physics, State University of New York at Binghamton, Binghamton, New York 13902-6000}

\author{Itsuko S. Suzuki}
\affiliation{Department of Physics, State University of New York at Binghamton, Binghamton, New York 13902-6000}

\author{Lingyan Wang}
\affiliation{Department of Chemistry, State University of New York at Binghamton, Binghamton, New York 13902-6000}

\author{Chuan-Jian Zhong}
\affiliation{Department of Chemistry, State University of New York at Binghamton, Binghamton, New York 13902-6000}


\date{\today}

\begin{abstract}
The aging and memory effects of Fe$_{3}$O$_{4}$ nanoparticles have been studied using a series of zero-field cooled (ZFC) and field-cooled (FC) magnetization measurements at various aging protocols. The genuine ZFC magnetization after the ZFC procedure with a single stop and wait process shows an aging dip at the stop temperature on reheating. The depth of the aging dip is dependent on the wait time. The frequency dependence of the AC magnetic susceptibility is indicative of critical slowing down at a freezing temperature $T_{f}$ ($= 30.6 \pm 1.6$ K). The relaxation time $\tau$ is described by a power law form with a dynamic critical exponent $x$ ($= 8.2 \pm 1.0$) and a microscopic relaxation time $\tau_{0}$ [$=(1.33 \pm 0.05) \times 10^{-9}$ sec]. The ZFC-peak temperature decreases with increasing magnetic field ($H$), forming a critical line with an exponent $p = 1.78 \pm 0.26$, close to the de Almeida-Thouless exponent ($p = 3/2$). These results indicate that the superspin glass phase occurs below $T_{f}$.
\end{abstract}


\maketitle



\section{\label{intro}Introduction}
The aging and memory effects of ferromagnetic nanoparticles have been the focus of extensive studies in recent years.\cite{REF01,REF02,REF03,REF04,REF05,REF06,REF07,REF08,REF09,REF10,REF11,REF12,REF13} Each ferromagnetic nanoparticle has a large magnetic moment (so-called superspin). Depending on the interactions between superspins, these systems are classed into two types. The noninteracting superspins give rise to superparamagnetic behavior. The superspins thermally fluctuate between their easy directions of magnetization and freeze along these directions at the blocking temperature $T_{b}$, where the relaxation time $\tau$ becomes equal to the measuring time $\tau_{m}$. Thus the superparamagnet (SPM) has a ferromagnetic blocked state below $T_{b}$. The relaxation time typically obeys an Arrhenius law. When the interactions between superspins, which are fully frustrated and random, become sufficiently strong, the interacting superspins cause spin frustration effect, resulting in superspin glass (SSG) behavior below a freezing temperature $T_{f}$. The low temperature spin-glass (SG) phase is experimentally characterized by observation of the flatness of the FC susceptibility below $T_{f}$, a critical slowing down of the relaxation time $\tau$ from the AC magnetic susceptibility, and a divergent behavior of the nonlinear susceptibility.\cite{REF03,REF07} The relaxation time $\tau$ which can be determined from the shift of the peak temperature of the AC magnetic susceptibility $\chi^{\prime}$ (dispersion) vs temperature ($T$) curve with frequency, exhibits a critical slowing down for SSG's.\cite{REF02,REF06}

The non-equilibrium properties of SSG's and SPM's have been observed in various nanoparticle systems, as a wait time dependence of zero-field cooled (ZFC) and field cooled (FC) magnetizations under various cooling protocols.\cite{REF01,REF02,REF04,REF05,REF06,REF08,REF09,REF10,REF11,REF12,REF13} The aging and memory effects of the SSG's are rather different from those of SPM's. A broad distribution of relaxation times characterize SPM's, while a crtitical slowing down occurs in the SSG's. The main features of their aging and memory effects are summarized as follows. These features provide a very unique method to determine dynamics governed by spin correlations between nanoparticles in SSG's and SPM's (Sasaki et al. \cite{REF12}).\\
(1)	(Genuine ZFC measurement). Only for SSG's, the ZFC magnetization $M_{ZFC}$ shows an aging dip at a stop temperature on reheating after the ZFC protocol with a single stop and wait process. The depth of the aging dip depends on the wait time $t_{w}$.\cite{REF08,REF10,REF12}\\
(2)	(Genuine FC measurement). For both SSG's and SPM's, the memory effect of $M_{FC}$ during a FC protocol with intermittent stop and wait processes are observed. A decrease of $M_{FC}$ is observed with decreasing $T$ for SSG's, while an increase of $M_{FC}$ is observed with decreasing $T$ for the SPM's.\cite{REF09,REF11,REF12,REF13}\\
(3)	(ZFC relaxation rate). Only for SSG's, the corresponding relaxation rate $S_{ZFC}(t, t_{w})$ [$=(1/H)dM_{ZFC}/d\ln t$] has a peak around $t = t_{w}$, as observed in spin glasses (aging effect).\cite{REF01,REF05,REF08}

In the present work, we have studied the magnetic properties of Fe$_{3}$O$_{4}$ nanoparticles. Synthesis and characterization of Fe$_{3}$O$_{4}$ nanoparticles used in the present work has been reported in detail previously.\cite{REF14,REF15} A simple review on the sample characterization will be presented in Sec.~\ref{exp}. We have measured the ZFC susceptibility ($\chi_{ZFC}$), FC susceptibility ($\chi_{FC}$) and AC magnetic susceptibility ($\chi^{\prime}$, $\chi^{\prime\prime}$) of Fe$_{3}$O$_{4}$ nanoparticles at various cooling protocols using a SQUID (superconducting quantum interference device) magnetometer. We show that the aging and memory effects, critical slowing down, and the flatness of the FC susceptibility at low temperatures, are clearly observed in Fe$_{3}$O$_{4}$ nanoparticles. These results indicate that the SSG phase occurs below a spin freezing temperature $T_{f}$ ($= 30.6 \pm 1.6$ K). 

The $H$-$T$ diagrams are examined from the temperature dependence of $\chi_{ZFC}$ and $\chi_{FC}$ of Fe$_{3}$O$_{4}$ nanoparticles at various $H$. The peak temperatures of the ZFC susceptibility of this system is determined as a function of $H$. We show that the ZFC-peak temperature $T_{p}$ ($=T_{f}$) for Fe$_{3}$O$_{4}$ nanoparticles decreases with increasing $H$, forming a critical line with an exponent $p = 1.78 \pm 0.26$, close to the de Almeida-Thouless (AT) exponent (= 3/2).\cite{REF25} This critical line is the phase boundary between the SPM and SSG phases. These results can be well described by the SSG model of interacting Fe$_{3}$O$_{4}$ nanoparticle systems. 

The contents of the present paper are as follows. In Sec.~\ref{exp} experimental procedure is presented, including the characterization of Fe$_{3}$O$_{4}$ nanoparticles. In Sec.~\ref{result} we present experimental results on the ZFC susceptibility, FC susceptibility, and AC susceptibility of our systems under various cooling protocols. In Sec.~\ref{dis}, the AT exponent $p$ will be discussed.

\section{\label{exp}Experimental procedure}
Fe$_{3}$O$_{4}$ nanoparticles capped with mixed monolayer of oleic acid and oleylamine were synthesized using a modified protocol.\cite{REF14,REF15}  Briefly, 0.71 g Fe(acac)$_{3}$ (2 mmol) was mixed with 2 mL oleic acid $\sim 6$ mmol), 2 mL oleylamine ($\sim 6$ mmol), and 2.58 g 1,2-hexadecanediol (10 mmol) in 20 mL phenyl ether under argon atmosphere with vigorous stirring. The solution was heated to 210 $^\circ$C and refluxed for 2 hrs. After cooling to room temperature, ethanol was added into the solution. A dark-brown precipitate (Fe$_{3}$O$_{4}$ nanoparticles) was separated by centrifuging, followed by washing with ethanol and drying with nitrogen. The sample for this study was used as synthesis.

Transmission electron microscopy (TEM) result for Fe$_{3}$O$_{4}$ nanoparticles shows the particles displayed high monodispersity in size ($52 \pm 5 \AA$) and well isolated, which is characteristic of the presence of an organic shell on the particle surface.\cite{REF14,REF15} In domains with densely-packed nanoparticles in the TEM image, we can determine the average edge-to-edge distance ($\sim 20\AA$), which was found to be quite close to the value expected for interdigitation of the alkyl chains in the interparticle shells.\cite{REF14} The X-ray Powder Diffraction (XRD) patterns show that the nanoparticles are highly crystalline materials.\cite{REF14,REF15} The crystalline features are reflected by the excellent matching of the diffraction peaks with that for standard spectra of Fe$_{3}$O$_{4}$. The Thermogravimetry Analysis (TGA) data for Fe$_{3}$O$_{4}$ nanoparticles capped with mixed monolayer revealed a mass loss of $\sim$32\% for the organic shell, so the mass percentage of Fe$_{3}$O$_{4}$ in the sample (filling factor) is about 68\%.\cite{REF14} The further detail of Fe$_{3}$O$_{4}$ nanoparticle synthesis and characterization were given in previous reports.\cite{REF14,REF15}

The DC magnetization and AC magnetic susceptibility were measured using a SQUID magnetometer (Quantum Design, MPMS XL-5). Before the measurements, a possible remnant magnetic field was removed using ultra low field option at 298 K: the resultant remnant field was less than 3 mOe. The measurements of the DC magnetization and AC magnetic susceptibility were carried out after appropriate cooling procedures. The details of the cooling protocol for each measurement are described in Sec.~\ref{result} and respective figure captions.

\section{\label{result}Experimental result}
\subsection{\label{sample}Curie-Weiss constant and saturation magnetization}
The DC magnetic susceptibility at $H = 1$ kOe was measured as a function of $T$ for $150 \le T \le 298$ K. It exhibits a well-defined Curie-Weiss behavior with a Curie-Weiss constant, $C_{g} = 11.39 \pm 0.30$ (emu K/g). The magnetization $M$ at both $T$ = 5.0 and 100 K was also measured as a function of $H$ for $0 \le H \le 45$ kOe. The magnetization $M$ saturates to a saturation magnetization $M_{s}$ (= 63.97 emu/g) above 20 kOe. The composition of the bulk Fe$_{3}$O$_{4}$ is described by FeO$\cdot$Fe$_{2}$O$_{3}$. The ferric (Fe$^{3+}$) ions are in a state with spin $S = 5/2$, while the ferrous (Fe$^{2+}$) ions are in a state with spin $S = 2$. The bulk Fe$_{3}$O$_{4}$ is a ferrimagnet with the Curie temperature 858 K. The spin magnetic moment of the Fe$^{3+}$ ions in the tetrahedral A sites are antiparallel to that in the octahedral B site. Then the magnetic moment of the Fe$^{3+}$ ions cancel out, leaving only the magnetic moment of the Fe$^{2+}$ ions in the octahedral B site.\cite{NEW01,NEW02} This means that there is one Fe$^{2+}$ mole per a molar mass $m_{0}$ (= 231.54 g) for one Fe$_{3}$O$_{4}$ mole. 

The average diameter of Fe$_{3}$O$_{4}$ nanoparticles is evaluated using using the data of the Curie-Weiss constant and the saturation magnetization in the following way. We assume that the mass ratio of Fe$_{3}$O$_{4}$ nanoparticles to the total sample is given by $f$. The parameter $f$ is the filling factor of Fe$_{3}$O$_{4}$ nanoparticles over the whole system\cite{REF14,REF15} and can be determined from the saturation magnetization $M_{s}$. The molar saturation magnetization $M_{s0}$ (emu/F$^{2+}$ mole) for the system is evaluated as
\begin{equation}
M_{s0}=(m_{0}M_{s})/f=1.481 \times 10^{4}/f  \text{  (emu/Fe}^{2+}\text{ mole)},
\label{eq1n}
\end{equation}
which is equal to the molar saturation magnetization $\bar{M}_{s}$ for Fe$^{2+}$ spins given by
\begin{equation}
\bar{M}_{s}=N_{A}g\mu_{B}S=2.368 \times 10^{4} \text{ (emu/Fe}^{2+}\text{ mole)}.
\label{eq2n}
\end{equation}
where $N_{A}$ is the Avogadro number, $\mu_{B}$ is the Bohr magneton, $g$ (= 2.12) is the Land\'{e} $g$-factor of Fe$^{2+}$ ion for the bulk Fe$_{3}$O$_{4}$,\cite{NEW03} and $S$ (= 2) is a spin of Fe$^{2+}$ ion. From Eqs.(\ref{eq1n}) and (\ref{eq2n}), the parameter $f$ can be estimated as $f = 0.625$. This value of $f$ is in good agreement with that determined from TGA ($f=0.68$).\cite{REF14} The molar Curie Weiss constant $C_{M}$ is given by
\begin{equation}
C_{M}=(m_{0}C_{g})/f=(2640 \pm 70)/f \text{ (emu K/Fe}^{2+}\text{ mole)}.
\label{eq3n}
\end{equation}
Then the average number $N_{0}$ of Fe$^{2+}$ atoms (which is also equal to the number of Fe$_{3}$O$_{4}$ molecules) in each nanoparticle can be estimated as 
\begin{equation}
N_{0} = (R/a_{0})^{3} = C_{M}/\bar{C}_{M} = (780 \pm 20)/f,
\label{eq4n}
\end{equation}
where $\bar{C}_{M}$ is the molar Curie-Weiss constant for the free Fe$^{2+}$ spins, $R$ is the average radius of Fe$_{3}$O$_{4}$ nanoparticles, $a_{0}$ ($= 2.63 \AA$) is the average radius of the sphere with the same volume occupied by one Fe$_{3}$O$_{4}$ molecule, and is defined by $a_{0} =[3m_{0}/(4\pi \rho N_{A})]^{1/3}$, $\rho$ (= 5.21 g/cm$^{3}$) is the density of bulk Fe$_{3}$O$_{4}$. The molar Curie-Weiss constant $\bar{C}_{M}$ for Fe$^{2+}$ ions is given by $\bar{C}_{M}= N_{A} \mu_{B}^{2}g^{2}S(S+1)/3k_{B}$ = 3.371 (emu /Fe$^{2+}$ mole K). From Eq.(\ref{eq4n}), the diameter $d$ of nanoparticles can be evaluated as 
\begin{equation}
d=2R=2a_{0}[(780 \pm 20)/f]^{1/3} .
\label{eq5n}
\end{equation}
The diameter $d$ can be estimated as $d = 56 \pm 5 \AA$ for $f = 0.625$ determined from the magnetization, and $d = 55 \pm 5 \AA$ for $f=0.680$ determined from TGA. These values are close to the ones obtained from the TEM micrograph; $d = 52 \pm 5 \AA$.\cite{REF14,REF15}

\subsection{\label{resultA}ZFC, FC and TRM magnetization}

\begin{figure}
\includegraphics[width=8.0cm]{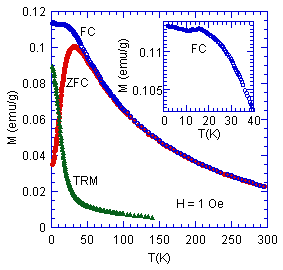}
\caption{\label{gra01}(Color online) $T$ dependence of $M_{ZFC}$, $M_{FC}$, and $M_{TRM}$ for Fe$_{3}$O$_{4}$ nanoparticles. $H$ = 1 Oe. The detail of the ZFC, FC, and TRM procedures is given in the text. The detail of $M_{ZFC}$ vs $T$ at $H=1$ Oe at low $T$ is shown in the inset.}
\end{figure}

Figure \ref{gra01} shows the $T$ dependence of the ZFC, FC, and thermoremnant (TRM) magnetization for Fe$_{3}$O$_{4}$ nanoparticles. These protocols used in the present work are explained as follows. (i) ZFC protocol: after the system was annealed at 298 K in the absence of $H$, it was cooled rapidly from 298 to 2.0 K. Immediately after the magnetic field $H$ (= 1 Oe) was turned on at 2.0 K, the ZFC magnetization was measured with increasing $T$ from 2.0 K to 298 K. (ii) FC protocol: After the system was annealed at 298 K in the presence of $H$, the FC magnetization was measured with decreasing $T$. (iii) TRM protocol: after the system was cooled from 298 to 2.0 K in the presence of $H$, the magnetic field was turned off. The TRM magnetization was then measured with increasing $T$ from 2.0 to 100 K in the absence of $H$. Note that in general the $T$ dependence of $M_{ZFC}$ is similar for SSG's and SPM's, while the $T$ dependence of $M_{FC}$ is noticeably different for the two. The FC magnetization $M_{FC}$ monotonically increases with decreasing $T$ for SPM's, while it tends to saturate to a constant value or even tends to decrease with decreasing $T$ for SSG's. In this sense, the $T$ dependence of $M_{FC}$ is a means for distinguishing between SPM's and SSG's. In the inset of Fig.~\ref{gra01}, we show the detail of $M_{FC}$ vs $T$ at $H=1$ Oe. Such a $T$ dependence of $M_{FC}$ is rather different from that of typical SSG. However, the slight decrease in $M_{FC}$ with decreasing $T$ below 16 K is indicative of the feature of $M_{FC}$ in the SSG. We note that the present system is not an ideal SSG system.

\begin{figure}
\includegraphics[width=8.0cm]{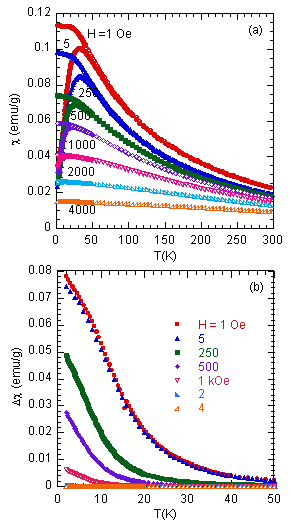}
\caption{\label{gra02}(Color online) $T$ dependence of (a) $\chi_{ZFC}$ and $\chi_{FC}$, and (b) $\Delta\chi$ ($= \chi_{FC} - \chi_{ZFC}$) of Fe$_{3}$O$_{4}$ nanoparticles. $H$ is changed as a parameter. 1 Oe $\le H\le 4$ kOe.}
\end{figure}

Figures \ref{gra02}(a) and (b) show the $T$ dependence of $\chi_{ZFC}$, $\chi_{FC}$, and $\Delta\chi$ [$= \chi_{FC} - \chi_{ZFC}$] for the Fe$_{3}$O$_{4}$ nanoparticles at various $H$. The susceptibility $\chi_{ZFC}$ at $H$ = 1 Oe shows a peak at $T_{p}$ ($\approx 32$ K) for $H$ = 1 Oe. The susceptibility $\chi_{FC}$ at $H$ ($\ge 5$ Oe) tends to saturate at low temperatures well below $T_{p}$. The difference $\Delta\chi$ gradually decreases with increasing $T$ and starts to appear at the onset temperature of irreversibility ($T_{irr}$). No sharp reduction of $\Delta\chi$ to zero is observed at $T = T_{irr}$, reflecting the volume distribution of Fe$_{3}$O$_{4}$ nanoparticles across the sample. Such a rounding effect of $T_{irr}$ in $\Delta\chi$ vs $T$ disappears at a higher $H$. The TEM measurement shows that the size distribution of Fe$_{3}$O$_{4}$ nanoparticles is similar to the log-normal distribution.\cite{REF14,REF15} Above $H$ = 500 Oe, $T_{irr}$ is very close to $T_{p}$. The flatness of $\chi_{FC}$ below $T_{p}$ and the coinciding of $T_{irr}$ and $T_{p}$ suggest that the Fe$_{3}$O$_{4}$ nanoparticles exhibit a SSG-like behavior. 

\subsection{\label{resultB}AC magnetic susceptibility}

\begin{figure}
\includegraphics[width=8.0cm]{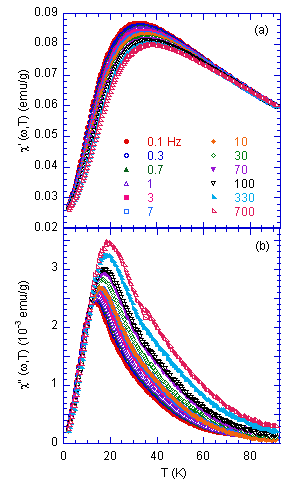}
\caption{\label{gra03}(Color online) $T$ dependence of (a) the dispersion $\chi^{\prime}$ and (b) the abosoprtion $\chi^{\prime\prime}$ for Fe$_{3}$O$_{4}$ nanoparticles. The frequency is changed as a parameter. $f$ = 0.1 - 1000 Hz. $h$ = 0.5 Oe. $T$ = 2 - 100 K. $H$ = 0.}
\end{figure}

\begin{figure}
\includegraphics[width=8.0cm]{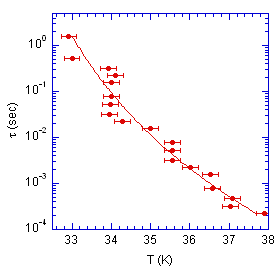}
\caption{\label{gra03c}(Color online) Relaxation time $\tau$ ($=1/(2\pi f$) vs $T$ at various frequencies $f$. $f$ = 0.1 - 1000 Hz. $T$ is equal to the peak temperature $T_{p}(\chi^{\prime})$ from $\chi^{\prime}$ vs $T$ curve. The solid line denotes a least-squares fit of the data of $\tau$ vs $T$ to the power law form given by Eq.(\ref{neq01}). The fitting parameters are given in the text.}
\end{figure}

Figures \ref{gra03}(a) and (b) show the $T$ dependence of the AC magnetic susceptibility: (a) the dispersion $\chi^{\prime}$ and (b) the absorption $\chi^{\prime\prime}$ at $H$ = 0. After the system was annealed at 298 K in the absence of $H$, it was rapidly cooled from 298 to 2.0 K. Both $\chi^{\prime}$ and $\chi^{\prime\prime}$ were measured at a fixed $T$ ($T\ge 2.0$ K) for various frequencies between 0.1 and 1000 Hz. After each measurement, the temperature was increased by $\Delta T$. The same measurement was then repeated at the temperature $T+\Delta T$. As shown in Fig.~\ref{gra03}(a), $\chi^{\prime}$ at $f = 0.1$ Hz shows a relatively broad peak at a peak temperature $T_{p}(\chi^{\prime})$ (= 32.5 K). This peak shifts to the high-$T$ side with increasing $f$: $T_{p}(\chi^{\prime})$ = 38.5 K for $f$ = 1 kHz. Also, the peak height of $\chi^{\prime}$ increases with increasing $f$. As shown in Fig.~\ref{gra03}(b), in contrast, the absorption $\chi^{\prime\prime}$ at $f$ = 0.1 Hz shows a relatively sharp peak at a peak temperature $T_{p}(\chi^{\prime\prime})$ (= 13.5 K). This peak shifts to the high-$T$ side with increasing $f$: $T_{p}(\chi^{\prime\prime})$ = 20 K for $f$ = 1 kHz. The peak height of $\chi^{\prime\prime}$ decreases with increasing $f$. It should be noted that $\chi^{\prime\prime}$ is independent of $f$ below 12 K. 

It is empirically known that the frequency shift in the peak temperature $T_{p}(\chi^{\prime})$ of $\chi^{\prime}$ vs $T$ curve, defined by $\Gamma =(1/T_{p})\Delta T_{p}/\Delta (\log_{10}\omega)$, offers a good criterion for distinguishing SG's ($\Gamma  < 0.06$) from SPM's ($\Gamma  \approx 0.3$).\cite{REF20} Our value of $\Gamma$ can be estimated as $\Gamma  \approx 0.05$, which suggests that our system is a SSG, and not a SPM. According to Hansen et. al,\cite{REF06} there are two criteria for the determination of the freezing temperature. First, the freezing temperature is defined as the temperature at which $\chi^{\prime\prime}$ attains 15\% of its maximum value. Second, the freezing temperature is defined from the relation $\chi^{\prime}(\omega,T_{f})=0.98\chi_{FC}(T=T_{f})$. Nevertheless, for convenience here we define the freezing temperature as the peak temperature $T_{p}(\chi^{\prime})$. Figure \ref{gra03c} shows the relaxation time $\tau$ which is estimated as $\tau = 1/(2\pi f)$ as a function of $T$ [= $T_{p}(\chi^{\prime})$]. The least-squares fit of the data of $\tau$ vs $T$ to a power law form for the critical slowing down,
\begin{equation}
\tau = \tau_{0}(T/T_{f}-1)^{-x} ,
\label{neq01}
\end{equation}
yields a dynamic critical exponent $x = 8.2 \pm 1.0$, a microscopic relaxation time $\tau_{0} = (1.33 \pm 0.5) \times 10^{-9}$ sec, and a freezing temperature $T_{f} = 30.6 \pm 1.6$ K. Our values of $x$ and $\tau_{0}$ are comparable with those of the Fe-C nanoparticles with a volume concentration 15 vol \% (superspin glass) reported by Hansen et al. \cite{REF06}: $x = 9.5$ and $\tau_{0} = 5.0 \times 10^{-9}$ sec. Note that our value of $x$ is also in good agreement with that of the 3D Ising spin glass Fe$_{0.5}$Mn$_{0.5}$TiO$_{3}$ ($x = 9.3 \pm 1.0$).\cite{REF21} These results indicate that our system is a SSG.

\subsection{\label{resultC}Memory effect in FC magnetization}

\begin{figure}
\includegraphics[width=8.0cm]{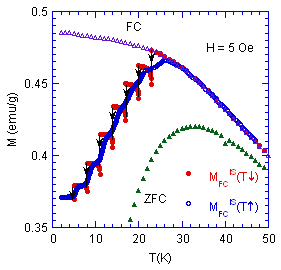}
\caption{\label{gra04}(Color online) $T$ dependence of $M^{IS}_{FC}(T\downarrow)$ ({\Large $\bullet$}) and $M^{IS}_{FC}(T\uparrow)$ ({\Large $\circ$}) for Fe$_{3}$O$_{4}$ nanoparticles, observed in the following FC aging protocol. The system is quenched from 298 to 50 K in the presence of $H$ (= 5 Oe). $M^{IS}_{FC}(T\downarrow)$ is measured with decreasing $T$ from 50 to 2.0 K but with intermittent stops (IS) at $T_{s}$ = 23, 20, 17, 14, 11, 8, and 5 K for a wait time $t_{s} = 1.0 \times 10^{4}$ sec. The field is cut off during each stop. The arrows indicate the relaxation of $M^{IS}_{FC}(T\downarrow)$. $M^{IS}_{FC}(T\uparrow)$ is measured at $H$ = 5 Oe with increasing $T$ after the above cooling process. The $T$ dependence of $M^{ref}_{FC}$ ($\triangle$) and $M^{ref}_{ZFC}$ ($\blacktriangledown$) are also shown as reference curves.}
\end{figure}

We present a peculiar memory effect observed in Fe$_{3}$O$_{4}$ nanoparticles  using a unique FC aging protocol. This effect also provides a good measure for determining whether the system is a SPM or a SSG.\cite{REF12} Figure \ref{gra04} shows the memory effect of the FC magnetization which is measured in the following way. First, the system was cooled using the FC protocol from 298 K to intermittent stop temperatures $T_{s}$ (= 23, 20, 17, 14, 11, 8, and 5 K) in the presence of $H$ (= 5 Oe). When the system was cooled down to each $T_{s}$, the field was turned off ($H = 0$) and the system was aged at $T_{s}$ for a wait time $t_{s}$ ($= 1.0 \times 10^{4}$ sec). The FC magnetization denoted by $M^{IS}_{FC}(T\downarrow)$ decreases with time $t$ due to the relaxation, where $IS$ stands for intermittent stop. After each wait time $t_{s}$ at $T_{s}$, the field ($H = 5$ Oe) was turned on and the cooling was resumed. We find that such an aging process leads to a step-like behavior of $M^{IS}_{FC}(T\downarrow)$ curve. Immediately after reaching 2.0 K, the magnetization $M^{IS}_{FC}(T\uparrow)$ was measured in the presence of $H$ (= 5 Oe) as the temperature was increased at a constant rate of 0.05 K/min. The magnetization $M^{IS}_{FC}(T\uparrow)$ thus measured exhibits step-like changes at each $T_{s}$. This implies that the spin configuration imprinted at each intermittent stop at $T_{s}$ for the wait time $t_{s}$ at $H = 0$ is retrieved by the curve on reheating. The magnetization $M^{IS}_{FC}(T\downarrow)$ is either parallel to $M^{ref}_{FC}$ as a reference at temperatures near $T_{s}$ = 23 and 20 K or is independent of $T$ at temperatures near $T_{s}$ = 14, 11, 8, and 5 K. The magnetization $M^{ref}_{FC}(T\downarrow)$ without intermittent stops is almost constant well below $T_{f}$ at $H=5$ Oe. The magnetization $M^{IS}_{FC}(T\downarrow)$ with intermittent stops decreases with decreasing $T$, while $M^{IS}_{FC}(T\uparrow)$ increases with increasing $T$. They meet together at temperatures a little above each stop temperature (approximately 1 K). Similar memory effects in the FC magnetization have been observed in the SSG Fe$_{3}$N nanoparticles.\cite{REF12} These features are in contrast to that of the SPM's such as ferritin (Sasaki et al.\cite{REF12} and Mamiya et al.\cite{REF04}), permalloy Ni$_{81}$Fe$_{19}$ (Sun et al.\cite{REF09}), Co particles (Zheng et al. \cite{REF11}): both $M^{IS}_{FC}(T\downarrow)$ with intermittent stops and $M^{ref}_{FC}(T\downarrow)$ without intermittent stops monotonically increase with decreasing $T$.

In summary, the decrease of $M^{IS}_{FC}(T\downarrow)$ with decreasing $T$ is a feature common to SSG's, while the increase of $M^{IS}_{FC}(T\downarrow)$ with decreasing $T$ is a feature common to SPM's.

\subsection{\label{resultD}Memory effect in ZFC susceptibility}

\begin{figure}
\includegraphics[width=8.0cm]{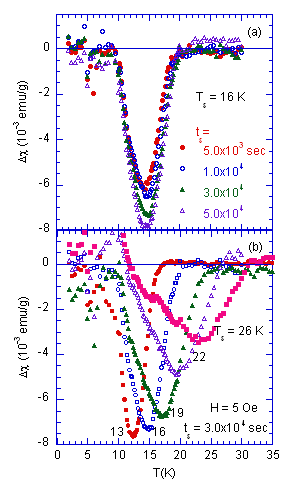}
\caption{\label{gra05}(Color online) (a) $T$ dependence of the genuine ZFC susceptibility for Fe$_{3}$O$_{4}$ nanoparticles. $\Delta\chi_{ZFC}=\chi^{SSW}_{ZFC}(T\uparrow)-\chi^{ref}_{ZFC}(T\uparrow)$. The system was annealed at $T$ = 100 K for 1200 sec. After the system was quickly cooled from 100 K to a stop temperature $T_{s}$ (= 16 K) at $H$ = 0, it was aged at $T_{s}$ for a wait time $t_{s}$ ($=5.0 \times 10^{4}$, $3.0 \times 10^{4}$, $1.0 \times 10^{4}$, $5.0 \times 10^{3}$, and $2.0 \times 10^{3}$ sec) [single stop and wait (SSW) process]. The cooling was resumed from $T_{s}$ to 2.0 K. Immediately after the field was turned on, the ZFC susceptibility $\chi^{SSW}_{ZFC}(T\uparrow)$ was measured at $H$ = 5 Oe with increasing $T$. The reference ZFC susceptibility $\chi^{ref}_{ZFC}(T\uparrow)$ was measured at $H$ = 5 Oe after the ZFC protocol without any stop and wait process. (b) $T$ dependence of $\Delta\chi_{ZFC}$. $t_{s} = 3.0 \times 10^{4}$ sec. $T_{s}=26$, 22, 19, 16, and 13 K. The ZFC protocol was the same as used in (a).}
\end{figure}

We measured the ZFC susceptibility of Fe$_{3}$O$_{4}$ nanoparticles after the ZFC aging protocol with a single-stop and wait (SSW) procedure. The sample was first rapidly cooled in zero-magnetic field from 100 K down to a stop temperature $T_{s}$. The system was aged at $T_{s}$ for a wait time $t_{s}$. The cooling was then resumed down to 2.0 K. Immediately after the magnetic field was turned on, the ZFC susceptibility $\chi^{SSW}_{ZFC}(T\uparrow)$ was measured on reheating. The reference ZFC susceptibility $\chi^{ref}_{ZFC}(T\uparrow)$ was also measured after the direct cooling of the system from 100 to 2.0 K without any stop and wait process. Figure \ref{gra05}(a) shows the $T$ dependence of the difference defined by $\Delta\chi_{ZFC}=\chi^{SSW}_{ZFC}(T\uparrow)-\chi^{ref}_{ZFC}(T\uparrow)$ for the SSW process, where $T_{s}$ = 16.0 K and $H = 5$ Oe. The wait times are chosen as $t_{s} = 5.0\times 10^{3}$, $1.0\times 10^{4}$, $3.0\times 10^{4}$, and $5.0\times 10^{4}$ sec, respectively. We find that the difference $\Delta\chi_{ZFC}$ takes a local minimum (an aging dip) at 15.9 K just below $T_{s}$. When the system is isothermally aged at $T_{s}$ = 16.0 K for $t_{s}$, its spin configuration gets arranged towards the equilibrium state. With further decrease in $T$, the equilibrated state becomes frozen in and the memory is retrieved on reheating. The depth of the aging dip is dependent on $t_{s}$, showing a clear evidence of the aging behavior that the domain size grows with time. We find here that the depth changes with increasing $t_{s}$ according to a power law form given by
\begin{equation}
\mid \Delta\chi^{SSW}_{ZFC} \mid_{dip}=At_{s}^{b} ,
\label{neq02}
\end{equation}
with $A = 0.0026 \pm 0.0002$ and $b = 0.10 \pm 0.01$. Similar time dependence of the aging dip has been observed in a 3D Ising SG Fe$_{0.5}$Mn$_{0.5}$TiO$_{3}$,\cite{REF23} where the depth of the aging dip logarithmically changes with $t_{s}$, rather than a power law form. We notice that our value of $b$ is nearly equal to the exponent $b^{\prime\prime}$ obtained from the time dependence of the absorption $\chi^{\prime\prime}(\omega,t)$ ($=A^{\prime\prime}t^{-b^{\prime\prime}}$) for Fe$_{0.5}$Mn$_{0.5}$TiO$_{3}$: $b^{\prime\prime} = 0.14 \pm 0.03$.\cite{REF24}

Figure \ref{gra05}(b) shows the $T$ dependence of the difference $\Delta\chi_{ZFC}$ at $H = 5$ Oe for the SSW process at $T_{s}$ (= 13, 16, 19, 22, and 26 K) for a wait time $t_{s}$ ($= 3.0 \times 10^{4}$ sec) during the ZFC protocol. The difference $\Delta\chi_{ZFC}$ clearly shows an aging dip. This dip occurs at the stop temperature $T_{s}$ where the system is aged during the SSW process. This result indicates the occurrence of the aging behavior. The depth of the aging dip is the largest at $T_{s}$ = 13.0 K and decreases with further increase in $T_{s}$. The width of the aging dip becomes broader as the stop temperature $T_{s}$ increases for $13 \le T_{s} \le 26$ K. Since the aging dip is expected to disappear for $T_{s}$ above $T_{f}$, this result indicates that the freezing temperature $T_{f}$ is at least higher than $T_{s}$ = 26 K. In fact, this result is consistent with our estimation of $T_{f}$ ($= 30.6 \pm 1.6$ K) which is derived in Sec.~\ref{resultB}. Similar $T$ dependence of the aging dip has been observed in Fe$_{0.5}$Mn$_{0.5}$TiO$_{3}$ \cite{RRR23} and canonical SG Ag (11 \% at. \% Mn).\cite{RRR23} Note that the detail of $T$ dependence of the peak vaue and width of the aging dip may be rather different for different systems.

\section{\label{dis}Discussion}

\begin{figure}
\includegraphics[width=8.0cm]{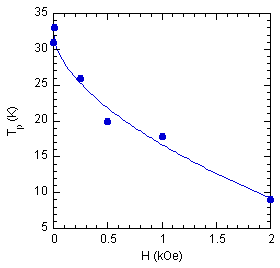}
\caption{\label{gra16}(Color online) Plot of $T_{p}$ as a function of $H$ for Fe$_{3}$O$_{4}$ nanoparticles. $T_{p}$ is a temperature at which $\chi_{ZFC} $ exhibits a peak. The solid line is least-squares fitting curve to Eq.~(\ref{eq24}) with $T_{g} = 32.5\pm 1.4$ K, $H_{0} = 3.61\pm 0.71$ kOe, and $p = 1.78\pm 0.36$.}
\end{figure}

In the mean-field picture, the phase transition of the SG systems can survive in the presence of low $H$, forming a critical line, the so-called de Almeida-Thouless (AT) line in the $H$-$T$ phase diagram \cite{REF25}
\begin{equation}
H(T)=H_{0} (1-T/T_{f} )^{p} ,
\label{eq24}
\end{equation}
where $T_{f}$ is the spin freezing temperature, $H_{0}$ is a field amplitude and the exponent $p = 3/2$. This line is the phase boundary between the PM (paramagnetic) phase and the SG phase. The correlation length and relaxation times diverge on crossing this line. In Fig.~\ref{gra16} we show the plot of the ZFC-peak temperature $T_{p}$ of $\chi_{ZFC}$ vs $T$ as a function of $H$ for Fe$_{3}$O$_{4}$ nanoparticles. The peak temperature $T_{p}$ decreases with increasing $H$. This critical line in the $H$-$T_{p}$ diagram may correspond to the phase boundary between the SPM and SSG phases. The least-squares fit of the data of $H$ vs $T_{p}$ for 1 Oe $\le H\le 2$ kOe to Eq.~(\ref{eq24}) yields the parameters $p = 1.78\pm 0.26$, $T_{f} = 32.5\pm 1.4$ K, and $H_{0} = 3.61\pm 0.71$ kOe. We find that $p$ is close to the AT exponent ($p$ = 3/2). These results indicate that there is an AT critical line in the $H$-$T$ phase diagram for Fe$_{3}$O$_{4}$ nanoparticles as a SSG system. Here we note that similar AT critical line has been reported by Sahoo et al. \cite{REF07} for ferromagnetic single domain particles of CoFe in discontinuous magnetic layers (Co$_{80}$Fe$_{20}$/Al$_{2}$O$_{3}$ multilayers). This system undergoes a SSG transition at a spin freezing temperature $T_{f}$. The peak temperature $T_{p}$ of the ZFC susceptibility shifts to the low-$T$ side with increasing $H$. The least-squares fit of the data of $T_{p}$ vs $H$ in the low-field range to Eq.~(\ref{eq24}) yields the exponent $p$ ($= 1.5\pm 0.4$), which is close to the AT exponent ($p$ = 3/2). In conclusion, the nature of the AT line in SSG systems is essentially the same as that in the SG systems.

The above discussion is based on the mean-field picture. The situation is rather different in the droplet picture.\cite{REF40} It is predicted that no phase transition occurs in the presence of even an infinitesimal $H$ as in the case of a ferromagnet. So there is no AT line in the $H$-$T$ phase diagram. Any apparent transition would be an artifact related to the limited experimental time scale. Several experimental results support the prediction from the droplet picture; the instability of the SG phase in thermal equilibrium in a finite $H$.\cite{REF41,REF42} 

\section{Conclusion}
The aging and memory effects of Fe$_{3}$O$_{4}$ nanoparticles have been studied in a series of DC magnetization measurements using various cooling protocols. The genuine FC magnetization after the FC procedure with multiple intermittent stop and wait processes shows a step-like increase at each stop temperature on reheating. The genuine ZFC magnetization after the ZFC procedure with a single intermittent stop and wait process shows an aging dip at the stop temperature on reheating. The depth of the aging dip is dependent on the wait time. The frequency dependence of the AC magnetic susceptibility for Fe$_{3}$O$_{4}$ nanoparticles is indicative of critical slowing down at a freezing temperature $T_{f}$ ($= 30.6 \pm 1.6$ K). The flatness of the FC susceptibility is observed below the ZFC-peak temperature $T_{p}$. The $H$ dependence of $T_{p}$ for Fe$_{3}$O$_{4}$ nanoparticles forms a critical line with an exponent $p = 1.78 \pm 0.26$, close to the de Almeida-Thouless exponent (= 3/2). These results are well described by the SSG model of interacting Fe$_{3}$O$_{4}$ nanoparticle system. 

\begin{acknowledgments}
The preparation of nanomaterials was supported by National Science Foundation (CHE 0349040).
\end{acknowledgments}

\end{document}